\let\oldvec\vec% Store \vec in \oldvec
\documentclass[oribibl,graybox]{svmult}
\let\vec\oldvec

\usepackage{graphicx}
\usepackage{subfigure}
\usepackage{algorithm2e}
\usepackage{amsmath}
\usepackage{balance}
\usepackage{multirow}
\usepackage{float}
\usepackage{stfloats}
\usepackage{flushend}
\usepackage{amssymb}
\usepackage{float}
\usepackage{mathptmx}       % selects Times Roman as basic font
\usepackage{helvet}         % selects Helvetica as sans-serif font
\usepackage{courier}        % selects Courier as typewriter font
\usepackage{type1cm}        % activate if the above 3 fonts are
\usepackage{makeidx}         % allows index generation
\usepackage{graphicx}        % standard LaTeX graphics tool
\usepackage{multicol}        % used for the two-column index
\usepackage[bottom]{footmisc}% places footnotes at page bottom
\makeindex     
\begin{document}
\title*{High Resilience Diverse Domain Multilevel Audio Watermarking with Adaptive Threshold}
\titlerunning{Multilevel Audio Watermarking}
%\author{*Jerrin Thomas Panachakel\thanks{*Jerrin Thomas Panachakel is with T\&BS SBU  at BEL, Bangalore, India (e-mail: jerrin.panachakel@gmail.com).}, \IEEEmembership{} and Anurenjan P.R., 
%\thanks{Anurenjan P.R.  is with the Dept. of Electronics and Communication Engineering at College of Engineering, Trivandrum, India (e-mail: anurenjan@ece.cet.ac.in).} \IEEEmembership{}}
\author{Jerrin Thomas Panachakel  \and Anurenjan P.R.}
\institute{Jerrin Thomas Panachakel\at
	Dept. of Electrical Engineering \\
	Indian Institute of Science\\
	Bangalore\\
	\email{jp@ee.iisc.ernet.in}           %  \\
	\and
	Anurenjan P.R.\at
	Dept. of Electronics and Communication Engineering\\
	College of Engineering\\
	Trivandrum\\
	\email{anurenjanpr@cet.ac.in
	}}

\maketitle
\abstract{A novel diverse domain (DCT-SVD \& DWT-SVD) watermarking scheme is proposed in this paper. Here, the watermark is embedded simultaneously onto the two domains. It is shown that an audio signal watermarked using this scheme has better subjective and objective quality when compared with other watermarking schemes. Also proposed are two novel watermark detection algorithms \textit{viz.,} \textit{AOT} (Adaptively Optimised Threshold) and \textit{AOTx} (\textit{AOT} eXtended). The fundamental idea behind both is finding an  optimum threshold for detecting a known character embedded along with the actual watermarks in a known location, with the constraint that the Bit Error Rate ($BER$) is minimum. This optimum threshold is used for detecting the other characters in the watermarks. This approach is shown to make the watermarking scheme less susceptible to various signal processing attacks, thus making the watermarks more robust.
}
%\begin{IEEEkeywords}
%Watermarking, discrete cosine transforms, discrete wavelet transforms
%\end{IEEEkeywords}
\section{Introduction}
{T}{he} earliest reference in literature about audio watermarking is a patent titled, ``Identification of sound and like signals,'' filed by Emil Hembrooke of the Muzac Corporation in the year 1954 \cite{1},\cite{2}. Ever since then, many watermarking algorithms have been developed and deployed. Interestingly, the field of application of watermarking itself witnessed a gradual change. Today, watermarking finds wider applications such as  transaction tracking, copyright protection, access control, broadcast monitoring etc. A good description of these applications can be found in \cite{5}. Recently, watermarking has been used for patient identification, tampering detection etc. in  wirelessly transmitted biomedical data such as ECG \cite{6}. All these applications requires the watermark to have one or more of the following properties,
\begin{enumerate}
\item{\textbf{Robustness:} }A watermark is said to be robust if it can survive common signal processing operations (commonly referred contextually as \textit{signal processing attacks}) such as compression, re-sampling etc. \cite{5}. A resourceful pirate can defeat a watermarking scheme by making it impossible to detect a watermark or making the scheme unreliable (i.e., a watermark is detected when no watermark is actually embedded) \cite{7} both of which should be prevented. Some applications do not require the watermark to be robust and the watermarking schemes used in these applications are referred to as \textit{fragile watermarking schemes} \cite{3}. 
\item{\textbf{Tamper Resistance:} }Tamper resistance refers to the watermark's ability to resist hostile attacks such as,
\begin{itemize}
\item \textit{Passive attacks} where the hacker tries to determine whether a watermark is present i.e. is trying to identify a covert communication.
\item \textit{Active attacks} where the hacker tries to remove the watermark.
\item \textit{Forgery attack} where the hacker tries to embed a new watermark \cite{11}.
\end{itemize}
Collusion attack is a special type of active attack where the hacker uses several copies of a work to produce a copy with no watermark \cite{8}.
\item \textbf{Fidelity:} A watermarking technique is said to have high fidelity if it causes only imperceptible degradation on the host signal. For instance, if the host  signal is an audio data, it should be difficult to distinguish by hearing the original audio and the audio signal with the watermark embedded. This criterion is often referred to as the \textit{perpetual transparency requirement}.
\item \textbf{Payload:} Payload refers to the amount of information that can be embedded into the given host signal. We require high payload while maintaining high perceptual transparency.
\item \textbf{Erasability:} In applications such as watermarking ECG signals in wireless transmission of biomedical signal, we require that the original host signal be recovered from the watermark signal.
\end{enumerate}
Many watermarking schemes can be found in literature. These schemes can be broadly classified into two groups,
\begin{enumerate}
\item Transform domain watermarking where the watermark is embedded into the coefficients of the host in a transformed domain, like Discrete Wavelet Transform (DWT) \cite{12}, \cite{13}, \cite{14}, \cite{15}, Discrete Cosine Transform (DCT), \cite{17}, \cite{18}, Discrete Fourier Transform (DFT) \cite{19}, \cite{20}, \cite{21}, \cite{22}, DWT-SVD (Singluar Value Decomposition) \cite{23}, etc., and
\item Spatial mode watermarking  where embedding is done in the spatial domain itself, like the schemes given in \cite{25} and \cite{26}.
\end{enumerate}
Based on the host signal, the watermarking can be audio watermarking, image watermarking, biomedical signal watermarking etc. 

Depending on the applications, the requirements on the watermarking scheme varies.
This paper proposes a watermarking scheme in which one watermark in embedded in the DWT-SVD domain and the other watermark in the DCT-SVD domain. The watermarking in the DWT-SVD domain follows \cite{23} but a novel watermarking scheme is followed in the DCT-SVD domain. Two different domains need to be used to avoid the interference between the two watermarks \cite{28}. Also, two novel generic watermark extraction algorithms \textit{viz. }, \textit{AOT} (Adaptively Optimised Threshold) and \textit{AOTx} (\textit{AOT} eXtended) are also proposed, which can be combined with any watermarking scheme that uses the ratio between a derived value (value obtained after applying transformation, decompositions, etc.) of the watermarked signal and the key (which is again a derived value but of the original host signal) as a threshold for extracting the watermark. It is shown that for the same payloads, the proposed diverse domain watermarking schemes outperforms conventional watermarking schemes in terms of imperceptibility. Also, it is shown that the use of \textit{AOT} and \textit{AOTx} at the detector side improves the robustness of the algorithm to various signal processing operations.

The rest of this paper is organised as follows, Section \ref{s2} gives an overview of the watermarking scheme; Subsections \ref{s4} and \ref{s5} discusses the embedding and extraction algorithms respectively. The performance metrics used for comparison are discussed briefly in Section \ref{sp}. Finally, the results of the comparison of the proposed watermarking scheme with other watermarking schemes in terms of Bit Error Rate ($BER$) when the transmitted signal is corrupted by varying amounts of Additive White Gaussian Noise (AWGN), Mean Opinion Score ($MOS$) criterion for different genres and $BER$ when subjected to various signal processing attacks are given in Section \ref{red}.
\section{Proposed Watermarking Scheme}
\label{s2}
\subsection{Overview}
Most of the multilevel watermarking algorithms found in literature are for either image watermarking \cite{27}, \cite{28} or for audio watermarking \cite{29}, \cite{29}, although a few works discuss about watermarking in other signals such as ECG signal \cite{6}. M. Butman\textit{ et al.} propose a three level watermarking scheme in \cite{28} in which spatial domain, DCT domain and wavelet domain are used for embedding watermarks in three levels. The primary purpose of this scheme is to detect tampering \cite{28}. Contrary to this approach, {S. Naveen \textit{et.al.}} \cite{29} use wavelet domain for multiple embedding with the primary purpose of increasing the payload. The primary disadvantage of the scheme by S. Naveen \textit{et al.} is that the secondary and further watermarks can be recovered only under ideal conditions (noise free conditions) \cite{29}. But in the proposed watermarking scheme, the primary and secondary watermarks can be recovered even under non-ideal conditions. The use of Singular Value Decomposition (SVD) improves the imperceptibility (transparency) and robustness of the watermark \cite{23}, \cite{29}.

In almost all non-blind or informed watermarking schemes \cite{5}, the watermark extraction involves the comparison of the derived value of the embedded signal received at the receiver with a key generated at the embedding stage. But due to noise or some signal processing attacks, the derived value of the embedded signal obtained at the receiver may vary from its actual value i.e., the value just before the transmission stage (it is assumed that the corruption with noise, signal processing attacks, etc. occur in the transmission stage). This alters the ratio between the derived value and the key  and this in turn will result in wrong decisions in the extraction stage. Adaptively Optimised Threshold (\textit{AOT}) and \textit{AOTx} (\textit{AOT} eXtended) aim to minimise, if not eliminate completely, the error caused in this manner. The fundamental idea behind both \textit{AOT} and \textit{AOTx} is to embed a known character at a known location in the host signal, along with the original watermarks. At the receiver, the threshold of the detector is optimised so that $BER$ of the detected watermark character at the location where the known character is embedded is minimum and this threshold is used for detecting the other characters in the watermarks. \textit{AOTx} is superior to \textit{AOT} in terms of the ability to determine the optimum threshold but has a cyclomatic complexity which is $6$ times higher than that of the latter, thus making it computationally infeasible to apply \textit{AOTx} at both levels.
\subsection{Embedding Algorithm}
\label{s4}\begin{figure}
\centering
\includegraphics[width=3.5in]{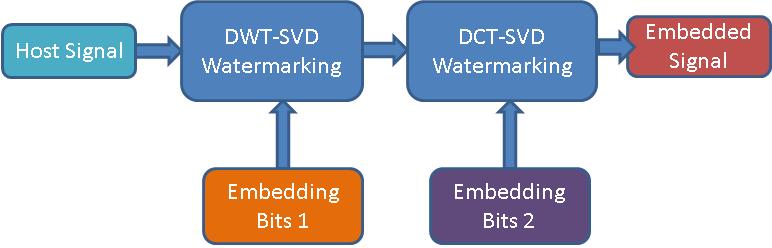}
\caption{HLD of multilevel watermarking scheme.}
\label{fig5}
\end{figure} 
The high level diagram of the embedding process of the proposed watermarking scheme is given in \mbox{Fig.\ref{fig5}.} Practically, there is only little significance in which watermarking is done first, whether it is the DCT-SVD or DWT-SVD, even though the watermarking is not strictly linear. In this work, DWT-SVD domain watermarking was done first followed by DCT-SVD domain watermarking. The two watermarking algorithms are discussed below.

\subsubsection{DWT-SVD Domain}
\label{dwt}
The watermarking scheme developed by Ali Al-Haj \textit{et al.} described in \cite{23} is used in this level, the only difference being that instead of binary image, the watermark is a text. Nevertheless, for the completeness of the discussion, the steps in the DWT-SVD watermarking scheme by Ali Al-Haj \textit{et al.} are given below,
\begin{itemize}
\item \textit{STEP 1: }The data to be watermarked, say $W$ is converted into its equivalent ASCII numbers and these ASCII numbers are converted to binary. Let the number of bits in the binary representation be $N$.
\item \textit{STEP 2: }The host audio signal is divided into $N$ frames. It is assumed that there are sufficient number of samples in the audio signal for this division. If the length of the audio signal, denoted as $L$ is not a multiple of $N$ sufficient zeroes are padded to the audio signal such that the new length  $L'$ is a multiple of $N$.
\item \textit{STEP 3: }Four level DWT is performed on each frame to obtain the details sub-bands $D_1$, $D_2$, $D_3$ and $D_4$ and approximate sub-band $A_4$.
\item \textit{STEP 4: }Arrange the details sub-bands as shown in \mbox{Fig. \ref{fig2}}. 
\begin{figure}
\centering
\includegraphics[width=2in]{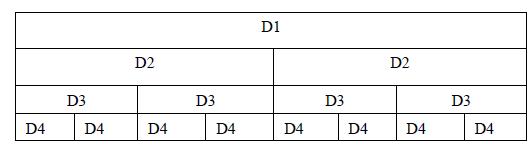}
\caption{Matrix formulation of X.}
\label{fig2}
\end{figure}
\item \textit{STEP 5: }Perform SVD operation on $X$ to obtain the orthogonal matrix $U,$ diagonal matrix $S$ and the unitary matrix $V,$ all of which are real  $4 \times 4$ matrices as given by the following equation,
\begin{equation}
SVD(X)= USV^T
\end{equation}
\item \textit{STEP 6: }Let the non-zero elements in the $n^{th}$ diagonal matrix $S_n$, corresponding to the $n^{th}$ frame be denoted as $S_{11}$, $S_{22}$, $S_{33}$ and $S_{44}$. Modify the value of the diagonal element $S_{11}$ as follows,
\begin{equation}
S_{11} = S_{11}\times (1 + \alpha \times w(n)) 
\end{equation}
\ \ where $\alpha$ is the watermark intensity, $w(n)$ is the $n^{th}$ watermark bit which is either $0$ or $1$. Higher the value of $\alpha$, more perceptible will the watermarking be. Lower values of $\alpha$ will make the detection more prone to errors. The original value of $S_{11}$, corresponding to each frame is stored as the ``key''. 
\item \textit{STEP 7: }Perform the inverse operations using the modified diagonal matrix to obtain the watermarked frame.
\item \textit{STEP 8: }Combine all frames to obtain the watermarked signal.
\end{itemize}
\begin{figure}
\centering
\includegraphics[width=3.5in]{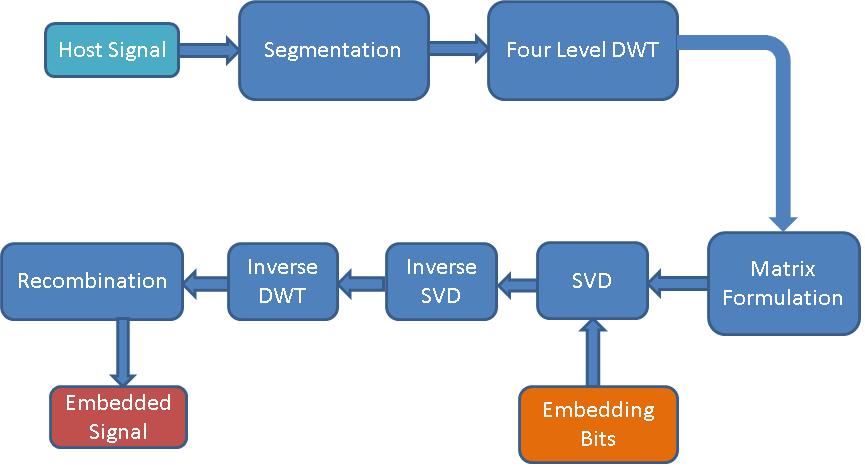}
\caption{HLD of DWT-SVD watermarking.}
\label{fig4}
\end{figure}
The high level diagram of DWT-SVD watermarking is given in \mbox{Fig. \ref{fig4}}.
\subsubsection{DCT-SVD Domain}
\label{dct}
The embedding in the DCT-SVD domain follows a novel approach and is performed as follows:
\begin{itemize}
\item \textit{STEP 1: }The data to be watermarked, say $W$ is converted into its equivalent ASCII numbers and these ASCII numbers are converted to binary. Let the number of bits in the binary representation be $N$.
\item \textit{STEP 2: }The host audio signal is divided into $N$ frames. It is assumed that there are sufficient number of samples in the audio signal for this division. If the length of the audio signal, denoted as $L$ in not a multiple of $N$, sufficient zeroes are padded to the audio signal such that the new length  $L'$ is a multiple of $N$.
\item \textit{STEP 3: }Apply DCT transformation on each frame.
\item \textit{STEP 4: }Form the matrix $X$ for each frame by using the first three, middle three and last three elements of each transformed frame as shown in Fig.\ref{fig1}.
\begin{figure}
\centering
\includegraphics[width=2.5in]{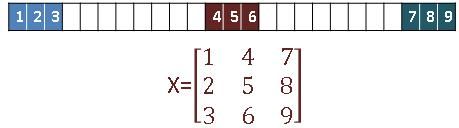}
\caption{Formation of matrix $X$ in DCT-SVD watermarking.}
\label{fig1}
\end{figure}
\item \textit{STEP 5: } Perform Singular Value Decomposition (SVD) operation on $X$ to obtain the orthogonal matrix $U,$ diagonal matrix $S,$ and the unitary matrix $V,$ all of which are real  $3 \times 3$ matrices as given by the following equation,
\begin{equation}
SVD(X)= USV^T
\end{equation}
\item \textit{STEP 6: }Let the non-zero elements in the $n^{th}$ diagonal matrix $S_n$ be denoted as $S_{11}$, $S_{22}$ and $S_{33}$. Modify the value of the diagonal element $S_{11}$ as follows,
\begin{equation}
S_{11} = S_{11}\times (1 + \alpha \times w(n)) 
\end{equation}
\ \ where $\alpha$ is the watermark intensity, $w(n)$ is the $n^{th}$ watermark bit which is either $0$ or $1$.
The original value of $S_{11}$ is stored as the ``key''. 
Let the modified $S$ matrix be denoted as $S'$.
\item \textit{STEP 7: }Obtain matrix $X'$ as,
\begin{equation}
X'=US'V^T
\end{equation}
\item \textit{STEP 8: }$X''$ is obtained by applying inverse DCT on $X'$. Apply the inverse of the operation performed in \textit{STEP 4} to obtain the watermarked frame.
\item \textit{STEP 7: }Combine the $N$ watermarked frames to obtain the watermarked audio signal.
\end{itemize}
\begin{figure}
\centering
\includegraphics[width=3.5in]{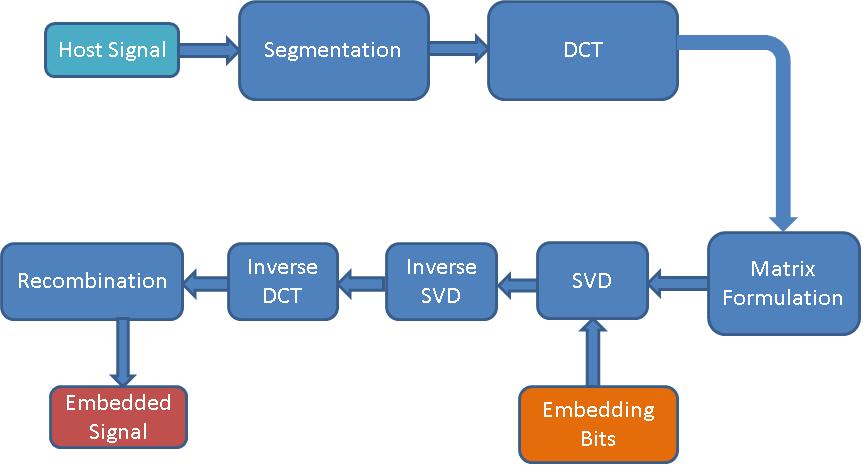}
\caption{HLD of DCT-SCD watermarking.}
\label{fig3}
\end{figure}
The high level diagram of DCT-SVD watermarking is given in \mbox{Fig. \ref{fig3}}.
\subsection{Extraction}
\label{s5}
For extraction, the inverse of the operations performed in the watermarking stage is performed. The watermark embedded last is extracted first and the one embedded first is extracted last. Though  it may not be a necessary step in many watermarking schemes, here we require the audio signal onto which the embedding was done should be recovered at least after the first extraction since this proves critical in extracting the next watermark with least error. 
The algorithm for extracting the watermark if a static threshold (threshold determined during the embedding stage) is used is as follows;

Performs \textit{STEPS 1-5} described in \mbox{Subsection \ref{dwt}} and in \mbox{Subsection \ref{dct}} for extracting the watermark embedded in the DWT-SVD domain and DCT-SVD domain respectively. Let the first element in the diagonal matrix obtained after applying Singluar Value Decomposition (SVD) be denoted as $S'_{11}$ and the same obtained in case of the original audio signal be denoted as $S_{11}$. It may be noted that it was these original values that were stored as the ``key'' in both domains. Theoretically, we detect the watermarked bit using the following rule,
\begin{eqnarray*}
&&if\ (\frac{S'_{11}}{S_{11}}\ ==\ th),\ then\ w(n)=1\\
&&else\ w(n)=0 
\end{eqnarray*}
\ \ where $w(n)$ is the $n^{th}$ watermarked binary bit and $th$ is the threshold which is equal to $(1+\alpha)$ where $\alpha$ is the watermark intensity.
Practically, we do not go for a stringent condition as above, instead, we give an allowance to accommodate errors by replacing the \textit{equality} condition with a \textit{comparative} condition like,
\begin{eqnarray*}
&&if\ (\frac{S'_{11}}{S_{11}}\ \ge\ th'),\ then\ w(n)=1\\
&&else\ w(n)=0 
\end{eqnarray*}
\ \ where $th'$ may be less than $th$ by a fraction of the value of $th$.

Now, when the transmitted signal gets modified due to some reasons such as addition of noise, signal processing operations, etc., the value of $S'{11}$ varies. Unless and otherwise a similar change in incorporated in either the ``key'' or the ``threshold'', the detections in the receiver side can go wrong. \textit{AOT} and \textit{AOTx} modifies the ``threshold'' to what we refer to as the \textit{optimum threshold} so that the detection errors are minimised.

 \textit{AOT} is used for DCT-SVD watermark extraction and \textit{AOTx} is used for DWT-SVD watermark extraction. Once the optimum threshold is determined for each domain, these optimum thresholds are used for detecting the other characters embedded in the host signal in the respective domains. How \textit{AOT} and \textit{AOTx} are used for obtaining the optimum thresholds is as follows.
\subsubsection{Adaptively Optimised Threshold (AOT)}
Both \textit{AOT} and \textit{AOTx} are used for determining the optimum threshold that will minimise the $BER$ at the detector. \textit{AOT} and \textit{AOTx} requires a known character to be embedded in a known location. The character ``U'' is chosen as the ``known character'' because the ASCII of ``U'' is $85$, which in binary notation is $1010101$, i.e., it has `1's and `0's in alternate locations, making the software programming easier. Also, it is the best go for a character having equal number of `1's and `0's in the binary of its ASCII equivalent.  

Consider an audio signal in which a watermark in embedded along with the character ``U'' which in embedded as the first character in the watermark. Perform \textit{STEPS 1 to 5} given in Subsection \ref{dct}. Let the non-zero elements in the $n^{th}$ diagonal matrix $S_n$ be denoted as $S_{11}$, $S_{22}$ and $S_{33}$. Also, let the $n^{th}$ watermark bit be denoted as $w(n)$. Note that in this case, $w(0)=1$, $w(1)=0$, $w(2)=1$, $w(3)=0$, $w(4)=1$, $w(5)=0$ and $w(6)=1$. Let $k(n)$ denote the original value of the top left element in the diagonal matrix obtained by performing SVD on the DCT applied frame, which can be obtained from the ``key''. We define the following  $6$ thresholds,
\begin{eqnarray}
thd_0=w(0)/k(0)\\
thd_1=w(1)/k(1)\\
thd_2=w(2)/k(2)\\
thd_3=w(3)/k(3)\\
thd_4=w(4)/k(4)\\
thd_5=w(5)/k(5)\\
thd_6=w(6)/k(6)
\end{eqnarray}
Note that $thd_0$, $thd_2$, $thd_4$ and $thd_6$ correspond to bit $1$ and $thd_1$, $thd_3$ and $thd_4$ corresponds to bit $0$. The optimum threshold, $th_{opt}$ is given by, 
\begin{equation}
th_{opt}=max(th0)+\frac{min(th1)-max(th0)}{2}
\end{equation} 
\ \ where $th0$ is a vector formed using $thd_1$, $thd_3$ \& $thd_5$ and $th1$ is a vector formed using $thd_0$, $thd_2$, $thd_4$ and $thd_6$.
\begin{figure}
\centering
\includegraphics[width=2.8in]{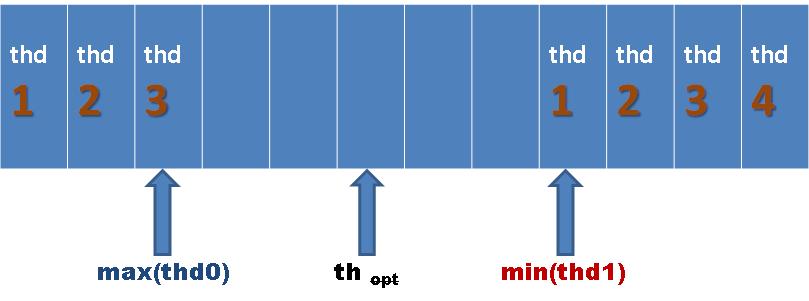}
\caption{Optimum threshold in \textit{AOT}.}
\label{fig6}
\end{figure}
Determination of optimum threshold $th_{opt}$ is shown graphically in \mbox{Fig. \ref{fig6}}. 
\subsubsection{Adaptively Optimised Threshold eXtended \textit{(AOTx)}}
\textit{AOTx} has $3$ stages; the first stage is exactly similar to \textit{AOT}. The threshold obtained from this stage is denoted as $th$. We define two variables $ze$ and $oe$. $ze$ is the ratio of number of `1's detected as `0's to the total number of `1's and $oe$ is the ratio of number of `0's detected as `1's to the total number of `0's. Clearly, the $BER$ is a linear combination of $ze$ and $oe$. For the character ``U'', $ze$, $oe$ and $SNR$ are related as,
\begin{equation}
SNR=\frac{oe\times 4+ze\times 3 }{7}
\end{equation}
 What we are trying to achieve in \textit{AOTx} is to minimise the sum of $ze$ and $oe$; i.e. we are actually minimising the $BER$.

If $ze$ is greater than $oe$, it means that the threshold is set at a lower value and therefore should be raised. Similarly, if $oe$ is greater than $ze$, it means that the threshold is set at a higher value and therefore should be lowered. If both $ze$ and $oe$ are $0$, it means that $th$ is the optimum threshold and there is no need for further optimisation. 

In the actual implementation, the updating rule is as follows:
\begin{eqnarray*}
&&if\ (ze>oe),\ then\ th=th+\frac{th}{2}\\
&&if\ (ze ==0\ and\ oe==0 ),\ then\ th_{opt}=th;\  flag=1;\\
&&else\ th=th-\frac{th}{2}
\end{eqnarray*} 
\begin{figure}
\centering
\includegraphics[width=2.5in]{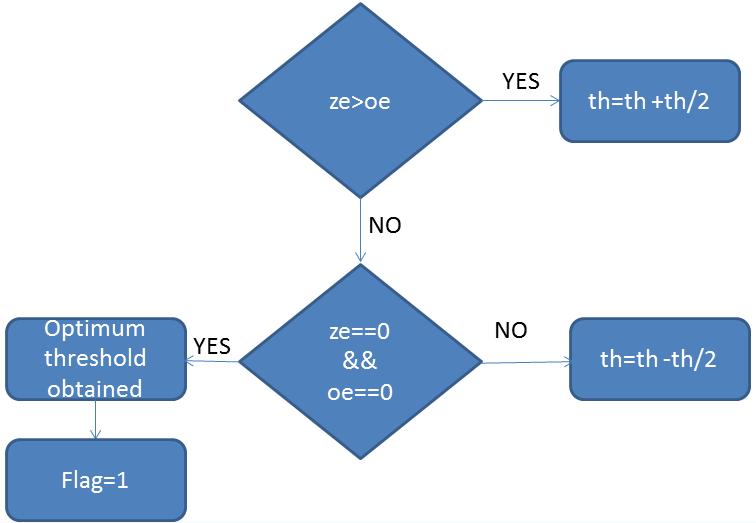}
\caption{\textit{AOTx}: Level 2.}
\label{fig7}
\end{figure} 
This is shown graphically in \mbox{Fig. \ref{fig7}}.
$flag$ is set when the optimum threshold is identified so that the final level of optimisation can be skipped to improve the computational time. In level 2, the change in threshold is rather coarse and during the implementation, we encountered situations where the thresholds determined in each subsequent iterations where oscillating between two values and the actual optimum threshold was between these two values. It is this observation that motivated us to design a third level in \textit{AOTx}.

If the $flag$ is set in level 2, level 3 is skipped. If it isn't set, the updating rule is similar to that of level 2 except that instead of having a coarse variation, we now have a fine variation as shown below, 
\begin{eqnarray*}
&&if\ (flag\ =\ 0 ),\ then\ th{opt}=th;\ break\\
&&if\ (ze>oe),\ then\ th=th+\frac{th}{10}\\
&&else\ th=th-\frac{th}{10}
\end{eqnarray*}
\section{Performance Metrics}
\label{sp}
The performance of the proposed watermarking scheme is compared with other watermarking techniques \textit{viz.} DWT-SVD watermarking \cite{23}, simple multilevel watermarking with static threshold, multilevel watermarking with \textit{AOT} in DCT-SVD domain and \textit{AOTx} in DWT-SVD domain, double embedding (multilevel) watermarking proposed by S. Naveen \textit{et al.} in \cite{29} (referred to hereafter as ``Double Embedding'') and DCT-SVD domain watermarking with static threshold (referred to hereafter as ``DCT-SVD based watermarking''). The static thresholds are decided at the watermark embedding stage itself. It may be noted that the DWT-SVD domain watermarking proposed by Ali Al-Haj \textit{et al.} has been shown in \cite{23} to surpass the STFT-SVD based watermarking proposed by H. Ozer. \textit{et al.} \cite{30} and DCT based watermarking \cite{31} proposed by I. Cox \textit{et al.}.

 Three criteria are used for performance analysis,
\begin{table}
\centering
\label{tab1}
\caption{MOS Grading Scale}
\begin{tabular}{|c|c|}
\hline
\textbf{MOS} & \textbf{Description} \\ 
\hline \hline 5 & Imperceptible \\ 
\hline 4 & Perceptible but not annoying \\ 
\hline 3 & Slightly annoying \\ 
\hline 2 & Annoying \\ 
\hline 1 & Very annoying \\ 
\hline 
\end{tabular}  
\end{table}
\begin{enumerate}
\item \textbf{Mean Opinion Score ($MOS$): }$MOS$ gives a numerical value for the perceived quality of a signal. $MOS$ was used in this work to analyse the perceivable difference of the audio quality, i.e. it is a measure of the transparency of the watermarking algorithm. $5$ scale $MOS$ is usually used with $5$ being the best value and $1$ being the worst value. The $MOS$ grading scale is given in \mbox{Table I}. Conventionally, $MOS$ is calculated by conducting listening test \cite{23}. Contrary to this, we used ``AQuA - Audio Quality Analyzer'' developed by\textit{ Sevana Oy} (a Finnish limited company founded in 2003, providing software development and services including a voice quality assessment software) for performing audio quality tests.
\item \textbf{BIT Error Rate ($BER$): }$BER$ was used to evaluate the watermark detection accuracy when the host signal is subjected to signal processing attacks, corrupted by noise etc. BER is defined as \cite{32}, \cite{23},
\begin{equation}
BER (W,\tilde{W})= \frac{\sum\limits_{i=1}^{siz}W(i)\oplus \tilde{W}(i)}{siz}
\end{equation}
For the watermarking schemes that doesn't support two simultaneous watermarks, the two watermarks are concatenated and $W$ denotes the concatenated watermark and $\tilde{W}$ denotes the watermark recovered at the detector. $siz$ corresponds to the size of the concatenated watermark, i.e. $W$. 
For the watermarking schemes that support multiple watermarks, $BER$ corresponding to both watermarks are calculated using the method described above and the average of the two is taken as the actual $BER$. 
\item \textbf{Signal to Noise Ratio ($SNR$): } $SNR$ is used to measure the objective quality of the \mbox{}{watermarked signal \cite{33}}. According to the recommendations of International Federation on Phonographic Industry (IFPI), the $SNR$  of a watermarked audio signal should be at least $20dB$.

$SNR$ is formulated in \cite{32} as:
\begin{equation}
SNR (X,\tilde{X})=10\log_{10} \frac{\sum\limits_{l=1}^{L}X(i)^2}{\sum\limits_{l=1}^{L}(X(i)-\tilde{X}(i)^2)} dB
\end{equation}
\ \ where $L$ is the number of samples in the original host signal $X(i)$ and the watermarked signal $\tilde{X(i)}$.

\end{enumerate}
\section{Results}

\label{red}

\begin{figure}
\label{as}
\centering{\subfigure[]{\includegraphics[width=3in]{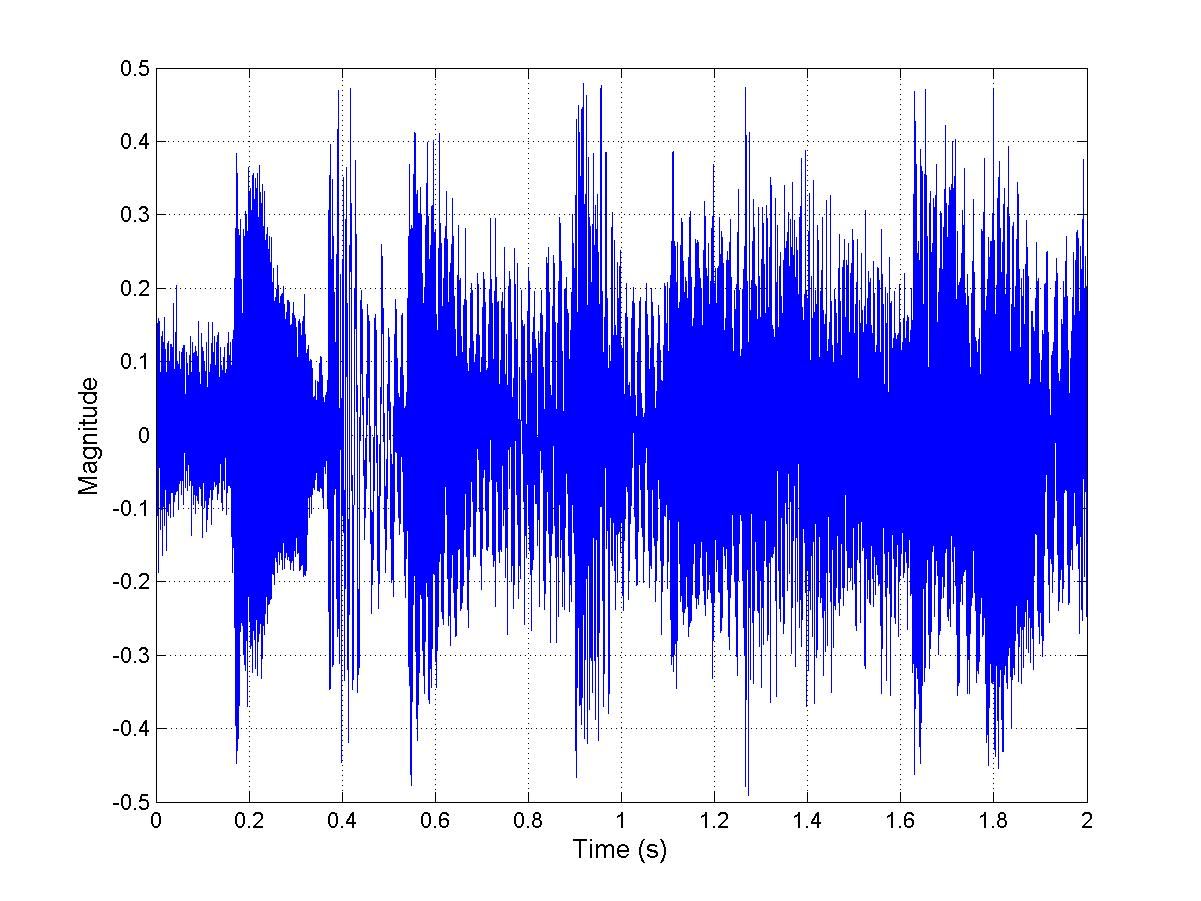}
\label{4a}}
\hfil\\
\subfigure[]{\includegraphics[width=3in]{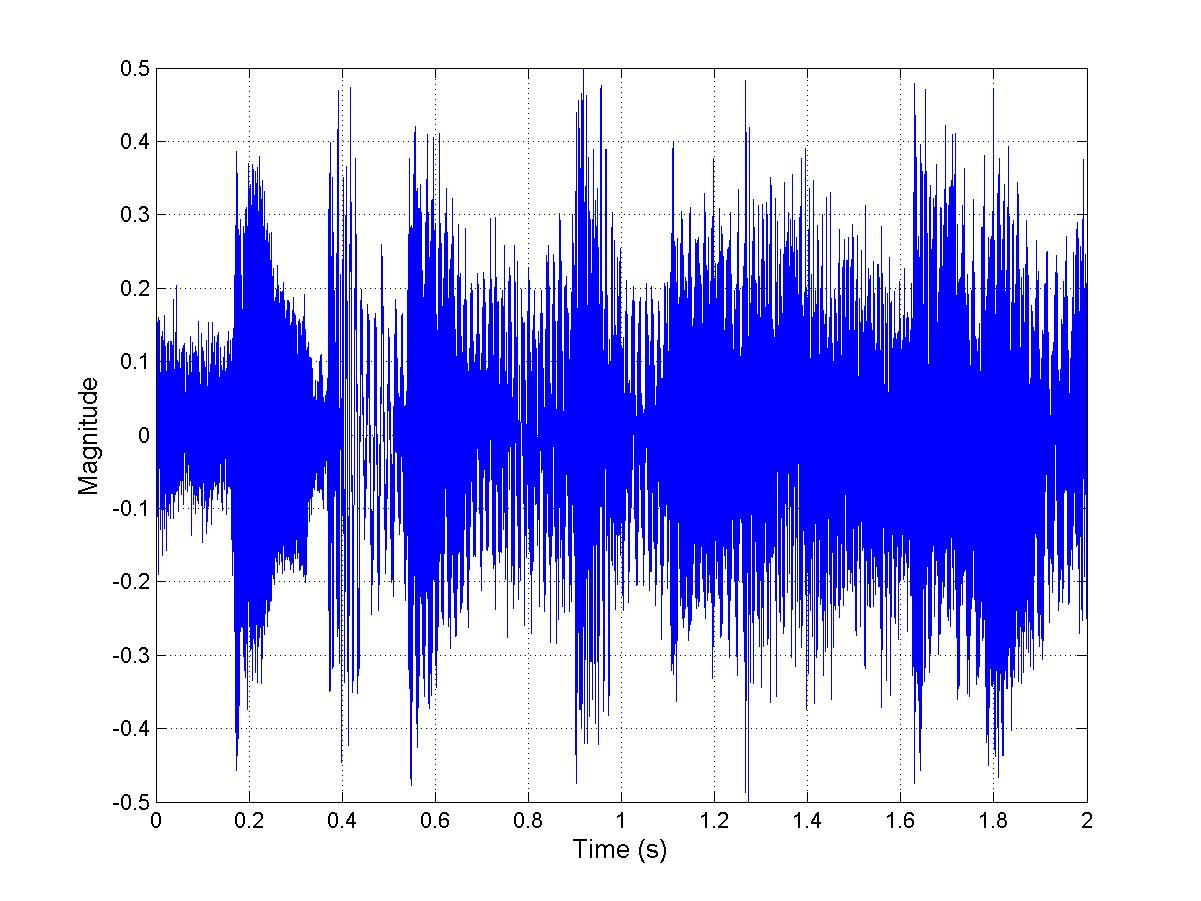}
\label{4b}}}
\hfil
\caption{(a) Waveform of original host audio signal (b) Waveform of the watermarked audio signal}
\end{figure}
\begin{table*}
\centering
\label{table}
\caption{$BER$ of various watermarking schemes when subjected to different signal processing attacks.}
	\scalebox{0.75}{
\begin{tabular}{|l|c|c|c|c|c|}

\hline \rule[-2ex]{0pt}{5.5ex} \textbf{Type of Attack \/ \textit{Watermarking Scheme}} & \textbf{\textit{Double Embedding}} & \textbf{\textit{Multilevel}}  & \textit{\textbf{Multilevel with AOT}} &\textbf{ \textit{DCT-SVD Domain}} &\textit{ \textbf{DWT-SVD Domain}}  \\ 
\hline
\hline \rule[-2ex]{0pt}{5.5ex} \textbf{Power Supply Hum} & \textbf{0.00} & \textbf{0.00} & \textbf{0.00} & 0.36 & 0.36 \\ 
\hline \rule[-2ex]{0pt}{5.5ex} \textbf{Amplification }& 0.21 & 0.42 & \textbf{0.00} & 0.42 & 0.42 \\ 
\hline \rule[-2ex]{0pt}{5.5ex} \textbf{Delay }& 0.21 & \textbf{0.14} & \textbf{0.14} & 0.42 & 0.42 \\ 
\hline \rule[-2ex]{0pt}{5.5ex} \textbf{Inversion} & \textbf{0.00} & \textbf{0.00} & \textbf{0.00} & \textbf{0.00} & \textbf{0.00} \\ 
\hline \rule[-2ex]{0pt}{5.5ex} \textbf{Linear Transform and Sparsification} & 0.35 & 0.14 & \textbf{0.00} & 0.35 & 0.50 \\ 
\hline 
\end{tabular} }
\end{table*}

The proposed watermarking scheme was simulated in \mbox{MATLAB\textsuperscript{\textregistered} 2012a (7.14)}. The waveform of the original host audio signal, which was of instrumental genre and the watermarked audio signal are given in \mbox{Fig.\ref{as}}. The watermark text in the DCT-SVD domain was \textit{``Jerrin Thomas Panachakel''} and that in the DWT-SVD domain was \textit{``College of Engineering, Trivandrum, India''}. 
%The ASCII values corresponding to the characters in the first watermark are  $74$
%   $101$
%   $114$
%   $114$
%   $105$
%   $110$
%    $32$
%    $84$
%   $104$
%   $111$
%   $109$
%    $97$
%    $32$
%    $80$
%    $97$
%   $110$
%    $97$
%    $99$
%   $104$
%    $97$
%   $107$
%   $101$
%   $108$.
%   Similarly, corresponding to the second watermark, the ASCII values are 
%       $67$
%      $111$
%      $108$
%      $108$
%      $101$
%      $103$
%      $101$
%       $32$
%      $111$
%      $102$
%       $32$
%       $69$
%      $110$
%      $103$
%      $105$
%      $110$
%      $101$
%      $101$
%      $114$
%      $105$
%      $110$
%      $103$
%       $44$
%       $32$
%       $84$
%      $114$
%      $105$
%      $118$
%       $97$
%      $110$
%      $100$
%      $114$
%      $117$
%      $109$
%       $44$
%       $32$
%       $73$
%      $110$
%      $100$
%      $105$
%       $97$.
       There were $24$ characters in the first watermark and the number of binary bits was $168$. Similarly, the number of characters in the second watermark was $41$ and the number of binary bits was $287$. The character 'U' was embedded in the first frame of both domains for facilitating the use of \textit{AOT} and \textit{AOTx}. The watermark intensity used was $\alpha=0.05$.
\begin{figure*}{h!}
\centering{\subfigure[]{\includegraphics[width=2.7in]{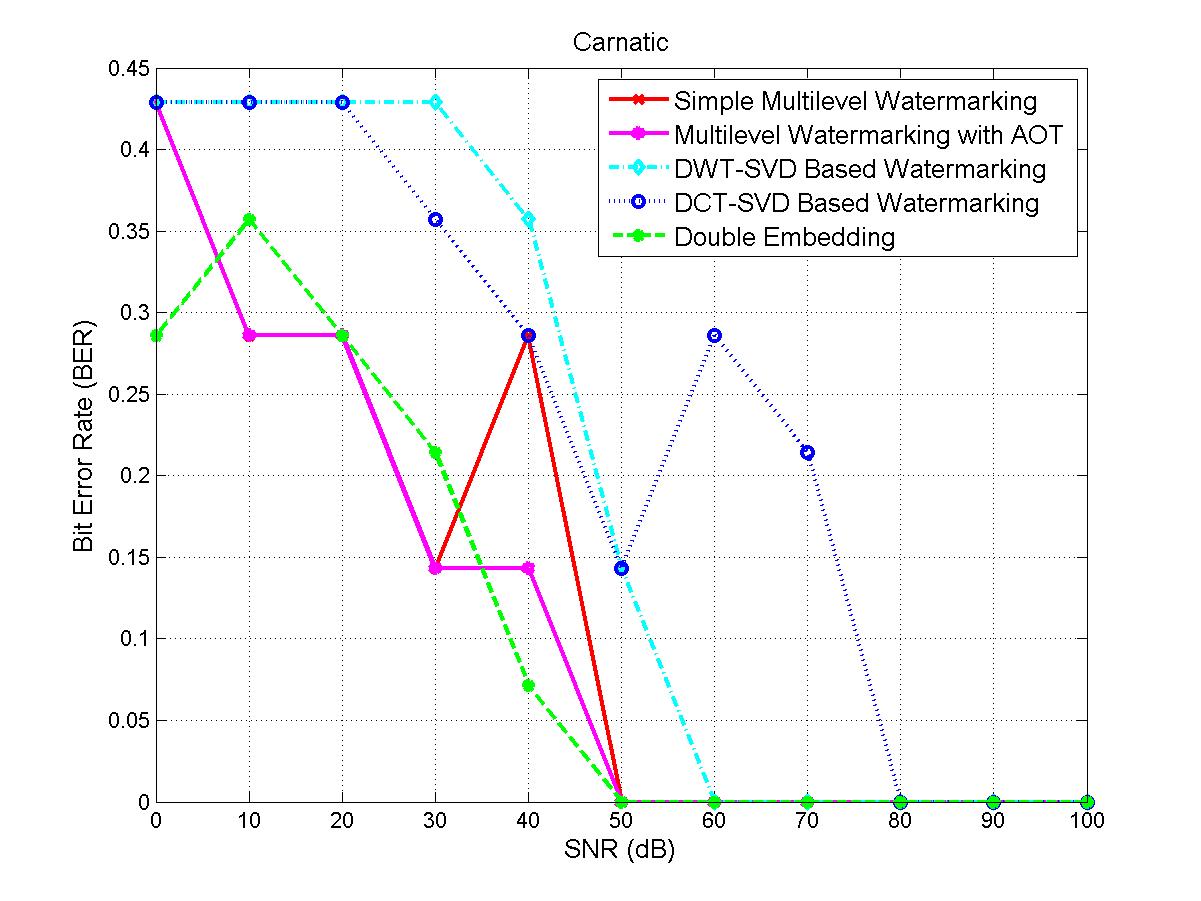}
\label{4akk}}
\subfigure[]{\includegraphics[width=2.7in]{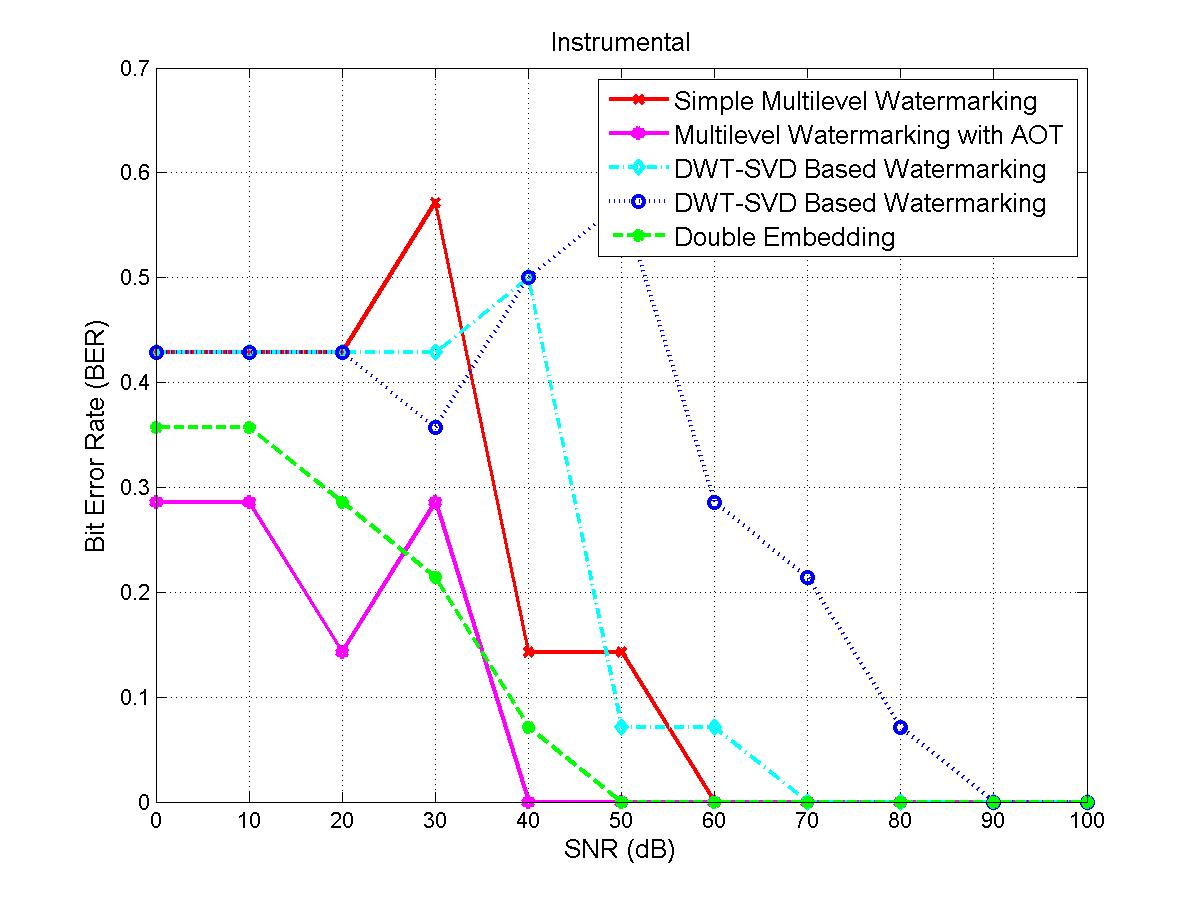}
\label{4bm}}}\\
\caption{$BER$ v/s $SNR$ for different genres. (a) Carnatic (b) Instrumental }
\label{f}
\end{figure*}
\begin{figure*}{h!}
	\centering{\subfigure[]{\includegraphics[width=2.7in]{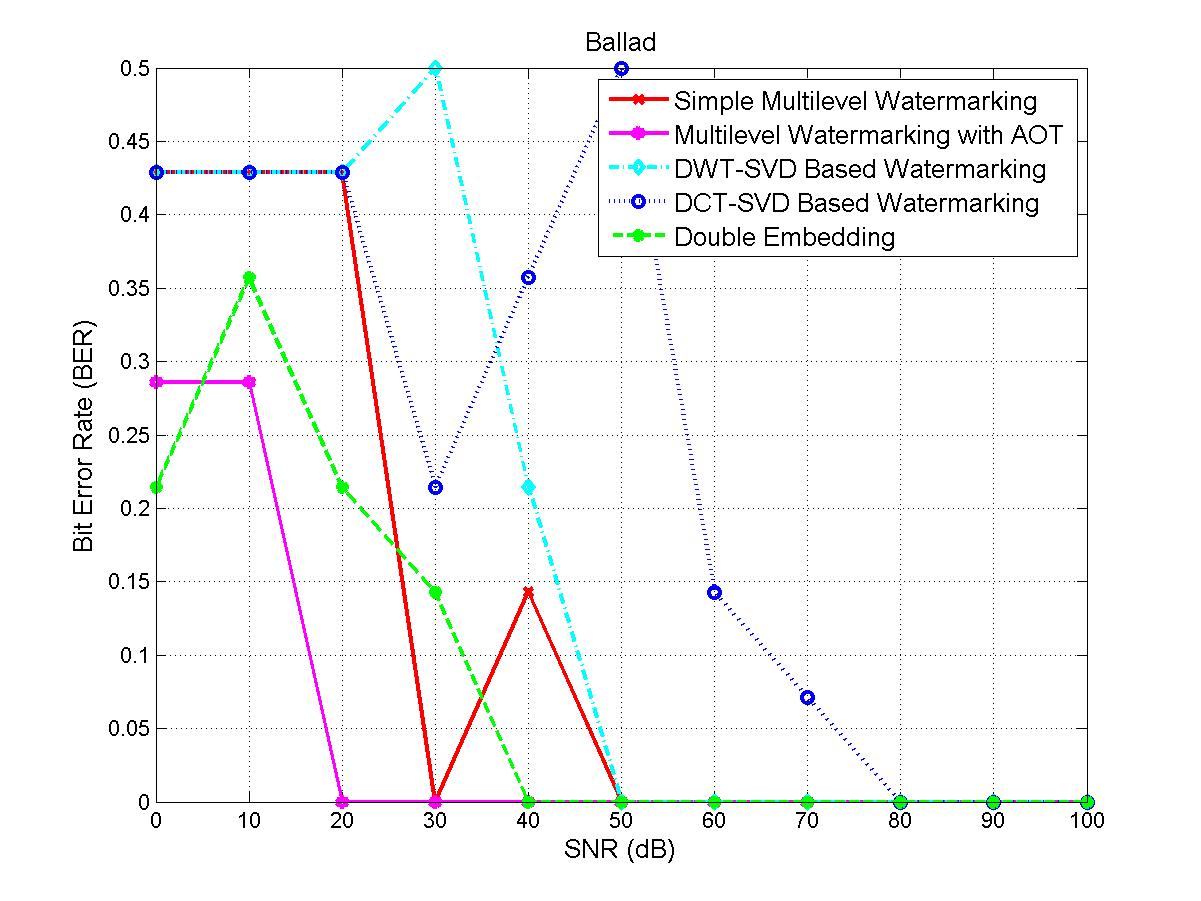}
			\label{4c}}
		\hfil
		\subfigure[]{\includegraphics[width=2.7in]{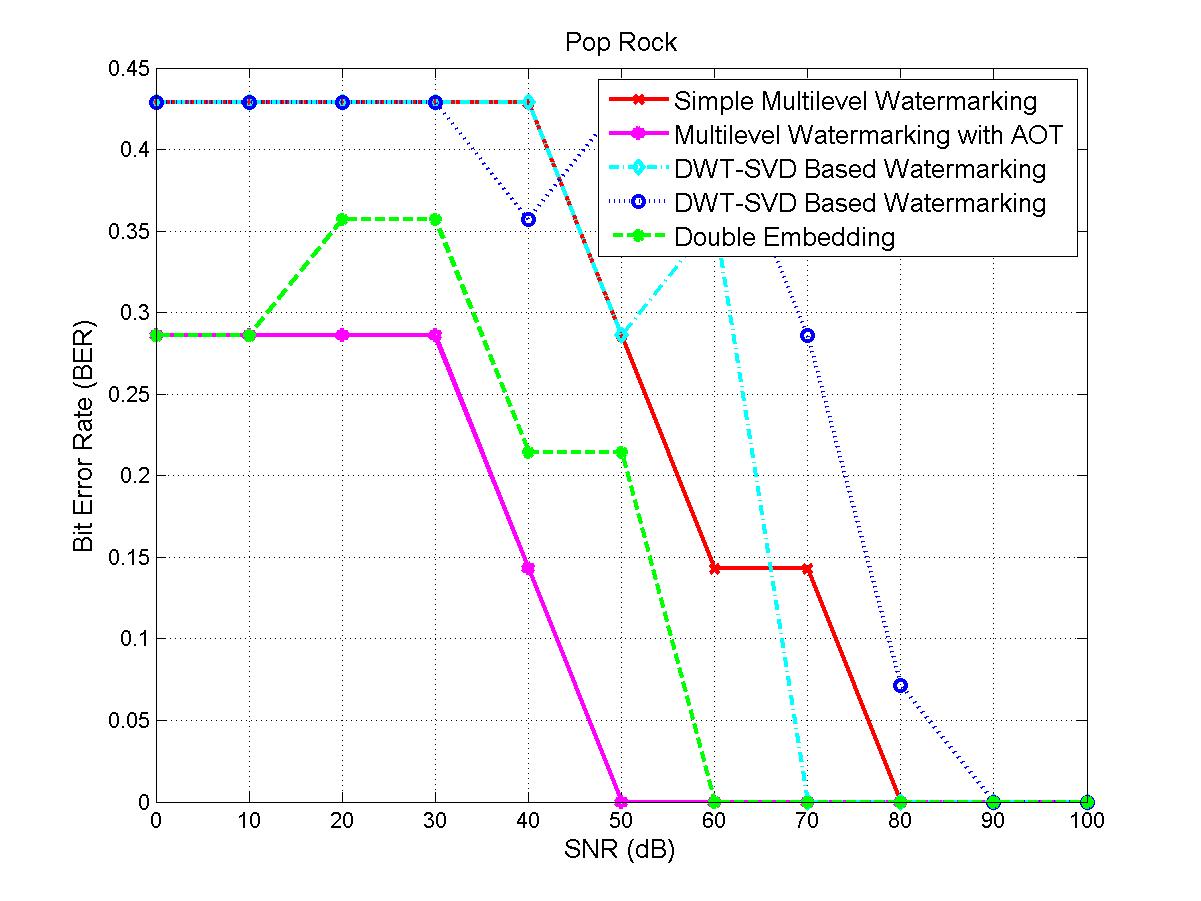}
			\label{4d}}
		\hfil
		\subfigure[]{\includegraphics[width=2.7in]{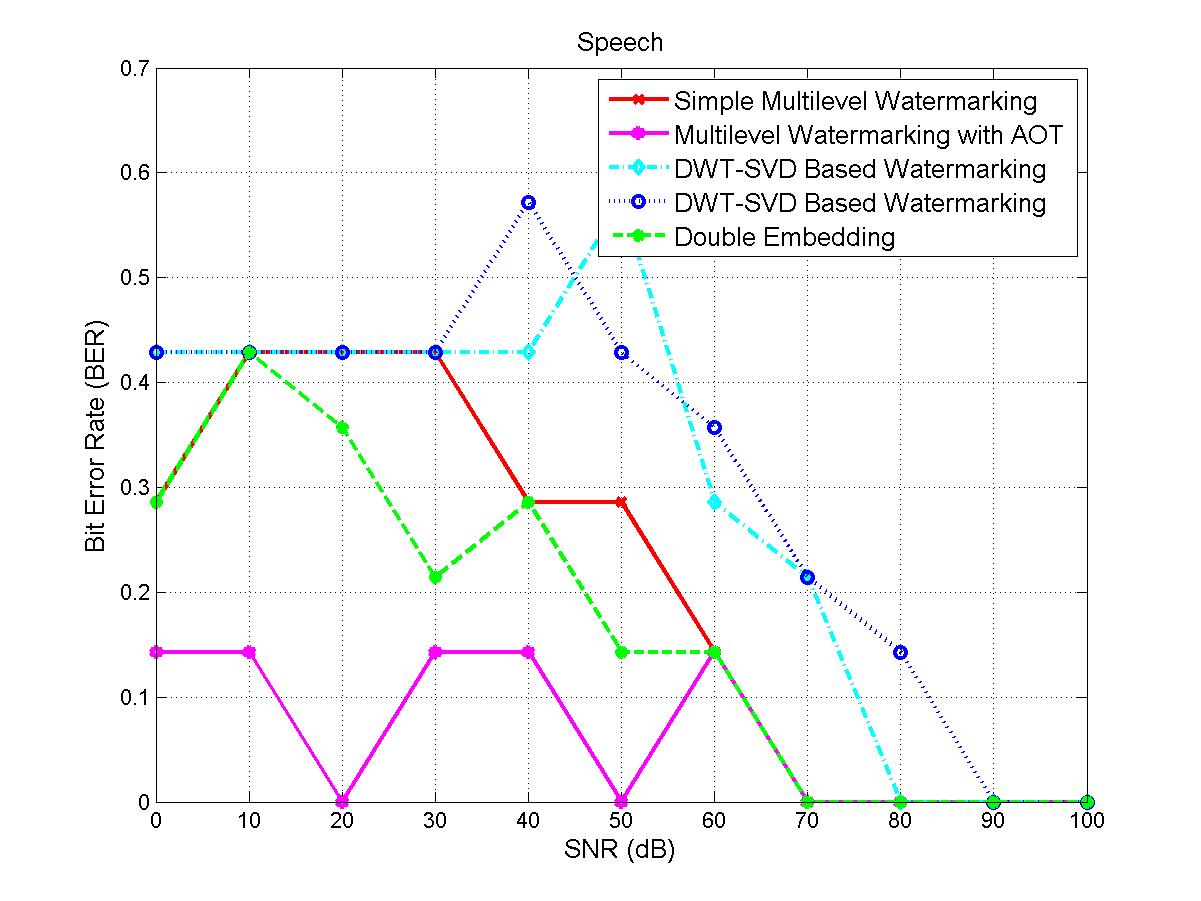}
			\label{4e}}}
	\hfil
	
	\caption{$BER$ v/s $SNR$ for different genres. (a) Ballad (b) Pop Rock (c) Speech}
	\label{ff}
\end{figure*}
\subsection{Bit Error Rate (BER)}
\subsubsection{AWGN Channel}
Watermarking was performed on several genres on audio signal and this audio signal was corrupted with Additive While Gaussian Noise (AWGN) with varying Signal to Noise Ratio ($SNR$). The $SNR$ was varied linearly from $0dB$ to $100dB$ in steps of $10dB$. This is assumed to model an AWGN channel through which the signal may pass during transmission. This assumption is supported by the central limit theorem and hence can be considered fair. Plots between $BER$ and $SNR$ for various genres are given in \mbox{Fig. \ref{f}. and Fig. \ref{ff}}. It is evident from the figure that Multilevel Watermarking with $AOT$ has the lowest $BER$ when the watermark signal is corrupted by AWGN noise. The plots are not monotonic in nature due to the randomness associated with the addition of noise.
\subsection{Signal processing attacks}
The host signal was watermarked using various watermarking schemes with two random watermarks if the scheme supports multiple watermarks or with the watermark obtained by concatenating the two random watermarks if the scheme supports only a single watermark. The watermarked signals were subjected to various signal processing attacks such as:
\begin{itemize}
\item \textbf{Power supply hum:} A $0.25V_{pp}$ sine wave of $50Hz$ is added to the watermarked audio signals to simulate the effect of addition of power supply hum. The corrupted watermarked signal was then processed for recovering the watermarks.
\item \textbf{Amplification: }Each of the watermarked signals is amplified by $14dB$.
\item \textbf{Delay: }The watermarked signals are delayed by $100ms$.
\item \textbf{Inversion: }All the samples in the watermarked signals are inverted.
\item \textbf{Linear Transformation and Sparsification: }Discrete Cosine Transform (DCT) was applied on the watermarked signals and the DCT coefficients whose absolute magnitude was less than $0.05$ was made zero. In our work, a speech signal was used as the host signal and $68\%$ of the DCT coefficients were less than $0.05$ and was hence made zero. These operations were done to model the effects of a lossy compression. 
\end{itemize}
After applying the above signal processing operations, the resultant watermarked signal was processed for recovering the watermark signal. Contrary to the approach by Naveen \textit{et al.} in \cite{29} where filtering was done prior to extraction, we attempted the recovery of the watermark without filtering the corrupted watermarked signal. The $BER$ between the original signal and the watermarked signal of various watermarking schemes when subjected to the above signal processing attacks is given in \mbox{Table II}. In all the cases, the proposed watermarking scheme has comparative, if not better performance, i.e. lowest $BER$. 
\subsection{Mean Opinion Score ($MOS$) and Signal to Noise Ratio ($SNR$)}
\begin{table}
\centering
\label{t3}
\caption{$MOS$ and $SNR$ values for $\alpha=0.05$ }
\begin{tabular}{|c|c|c|}
\hline \rule[-2ex]{0pt}{5.5ex} \textbf{Watermarking Technique} & \textbf{MOS} & \textbf{SNR (dB)}\\ \hline
\hline \rule[-2ex]{0pt}{5.5ex} Double Embdedding & 4.55 & 27.00 \\ 
\hline \rule[-2ex]{0pt}{5.5ex} Multilevel Watermarking & 4.58 & 33.20 \\ 
\hline \rule[-2ex]{0pt}{5.5ex} DCT-SVD  & 4.33 & 60.24 \\ 
\hline \rule[-2ex]{0pt}{5.5ex} DWT-SVD & 4.33 & 30.45\\ 
\hline 
\end{tabular} 
\end{table}
\begin{table}
\centering
\label{t4}
\caption{$MOS$ and $SNR$ values for $\alpha=0.1$ }
\begin{tabular}{|c|c|c|}
\hline \rule[-2ex]{0pt}{5.5ex} \textbf{Watermarking Technique} & \textbf{MOS} & \textbf{SNR (dB)}\\ \hline
\hline \rule[-2ex]{0pt}{5.5ex} Double Embdedding & 4.37 & 20.76 \\ 
\hline \rule[-2ex]{0pt}{5.5ex} Multilevel Watermarking & 4.55 & 27.18 \\ 
\hline \rule[-2ex]{0pt}{5.5ex} DCT-SVD  & 4.26 & 54.11 \\ 
\hline \rule[-2ex]{0pt}{5.5ex} DWT-SVD & 4.24 & 24.42\\ 
\hline 
\end{tabular} 
\end{table}
The $MOS$ and $SNR$ values were calculated for different watermark intensities $\alpha=0.05$ \& $\alpha=0.1$ for a speech signal sampled at $8KHz$ and is shown in \mbox{Table III \& IV}. Clearly, the proposed technique has better $MOS$ values, proving our argument that watermarking on two diverse domains has better subjective  quality than watermarking on a single domain. The values obtained for $SNR$ satisfies the International Federation on Phonographic Industry (IFPI) recommendations. It is interesting to note that although the DCT-SVD domain based watermarking is vulnerable to signal processing attacks, it has an $SNR$ greater than the value specified by IFPI for a watermarked audio signal, $60.24 dB$ and $54.11 dB$ for $\alpha$ values $0.05$ and $0.1$ respectively. These values are highest when compared to the values of other watermarking schemes discussed. Also, the $MOS$ values for DCT-SVD domain watermarking is comparable to multilevel watermarking.
\section{Conclusion}
Novel multilevel watermarking in diverse domains, DCT-SVD domain based watermarking and two generic watermark extraction algorithms \textit{viz.,} Adaptively Optimised Threshold \textit{(AOT)} and \textit{AOT} eXtended \textit{(AOTx)} are discussed.  The proposed watermarking algorithm has better subjective and objective quality, which is evident from the higher $SNR$ and $MOS$ values. The use of \textit{AOT} and \textit{AOTx} at the detection stage makes the watermarking schemes less susceptible to various signal processing attacks although the proposed technique fails for geometric signal processing attacks such as cropping, addition of delay etc. This demerit can be addressed and resolved by using synchronisation codes. The fact that \textit{AOT} and \textit{AOTx} are generic algorithms which can be incorporated in several existing watermarking schemes to improve their susceptibility towards signal processing attacks opens up a wide range of applications for both. Also, the developed DCT-SVD domain based watermarking has the highest $SNR$ eventhough it is more susceptible to signal processing attacks. Hence, this can be used as a ``fragile watermarking scheme''.
\section*{Acknowledgement}
We would like to place on record our deepest gratitude to 
\begin{itemize}
%\item Dr. Vrinda V. Nair, Dr. Jiji C.V., Dr. James T.A. \& Prof. Jeena R.S., faculty members and Ms. Revathy Sivanandan \& Ms. Lakshmi Mohan Vijaya, M.Tech. scholars of Dept. of Electronics and Communication Engg., College of Engineering, Trivandrum, India for their valuable suggestions and comments.
\item Mr. Valeri Sitnikov of ``Sevana Oy'' for providing us a free evaluation version of their product, ``AQuA - Audio Quality Analyzer'' for this work.
\end{itemize}
\bibliographystyle{splncs} 
\bibliography{b}
\end{document}